\documentclass[preprint,12pt]{elsarticle}

\usepackage{amsmath,amssymb}
\usepackage{booktabs}
\usepackage{array}
\usepackage{multirow}
\usepackage{graphicx}
\usepackage{float}
\usepackage{url}
\usepackage{xurl}
\usepackage{pdflscape}
\usepackage{ragged2e}
\usepackage{tabularx}
\usepackage{tikz}
\usetikzlibrary{arrows.meta,calc,positioning,shapes.geometric}
\biboptions{sort&compress}
\journal{Journal of Systems and Software}

\newcommand{\acceptstate}{\mathrm{Accept}}

\newcommand{\recoverable}{\mathrm{Recoverable}}
\newcommand{\successout}{\text{\texttt{success}}}
\newcommand{\blockedout}{\text{\texttt{blocked}}}
\newcommand{\failedout}{\text{\texttt{failed}}}
\newcommand{\skippedout}{\text{\texttt{skipped}}}
\newcommand{\acceptedflag}{\text{\texttt{accepted}}}
\newcommand{\withheldflag}{\text{\texttt{withheld}}}
\newcommand{\codeid}[1]{\texttt{\detokenize{#1}}}
\renewcommand{\path}[1]{\codeid{#1}}
\newcolumntype{L}[1]{>{\RaggedRight\arraybackslash}p{#1}}
\newcolumntype{C}[1]{>{\Centering\arraybackslash}p{#1}}
\newcolumntype{Y}{>{\RaggedRight\arraybackslash}X}

\begin{document}

\begin{frontmatter}

\title{Verify-Gated Completion as Admission Control in a Governed Multi-Agent Runtime: A Bounded Architecture Case Study}

\author[vimaru]{Hai-Duong Nguyen\corref{cor1}}
\ead{briannguyen290701@gmail.com}
\author[vimaru]{Xuan-The Tran}
\ead{thetx.vck@vimaru.edu.vn}
\cortext[cor1]{Corresponding author.}
\affiliation[vimaru]{organization={Vietnam Maritime University},
                     city={Haiphong},
                     country={Vietnam}}

\begin{abstract}
Multi-agent systems now coordinate tools, roles, and persistent state, but completion is still a weak point: a worker can claim that a task is done before the evidence is strong enough. This preprint reports a bounded architecture case study of verify-gated completion as admission control in a governed multi-agent runtime. It is not a benchmark or production-validation study. Agents may propose completion, while a read-only admission verifier decides whether the claim is admitted. Ambiguous cases resolve fail-closed, and packetized state plus event traces are retained for audit.

We ask: what can the released evidence support about auditable, verify-gated completion?

We examine one reference implementation through an architecture report and a bounded empirical snapshot. In the released verify-completed slice, the known-outcome invoked-event verify success share was 1,791/1,800 = 99.5\% with \texttt{outcome=success}. This is an accounting measure over invoked verification events, not a task-completion, production-reliability, or benchmark-success rate. Task-level verify coverage is not computable; 1,762/1,801 rows (97.83\%) came from one high-volume reporting cluster; and the slice is synthetic-heavy, with only 17 production-classified events. A shadow Policy/Governance Verifier (PGV) evaluation showed 1,526/1,548 = 98.58\% rule agreement, 0/1,526 false-success among safe-to-proceed predictions, and blocked precision of 2/518 = 0.39\%, so PGV remains advisory. The evidence supports architectural and evidentiary claims only: under the observed conditions, a read-only verify gate plus packetized admission records gave completion decisions an inspectable, fail-closed audit path. Claims about deployed operation, safety guarantees, outcome gains, task-level coverage, or external validity remain outside scope.
\end{abstract}

\begin{keyword}
case study\sep{} multi-agent systems\sep{} admission control\sep{} verification\sep{} auditability\sep{} traceability\sep{} software orchestration
\end{keyword}

\end{frontmatter}

\section{Introduction}\label{sec:intro}%

Large language model agents now appear in workflows that combine tools, specialized roles, and persistent state. Prior work such as ReAct, CAMEL, AutoGen, MetaGPT, Reflexion, Self-Refine, and SWE-agent shows how planning, tool use, role specialization, and iterative revision can improve agent behavior in research and software settings~\cite{react,camel,autogen,metagpt,reflexion,selfrefine,sweagent}. A more operational question is still less settled: when should the system accept that a task is complete?

Many agent papers describe how work is routed and produced. They are often less explicit about who may accept that work. The evidence required for acceptance, the handling of blocked states, and the record of stale context or rollback may remain implicit. When those conditions are loose, premature success claims are easier to surface and harder to audit later.

We treat this as a runtime-design problem rather than just a modeling problem. In the architecture examined here, execution may propose completion, but admission depends on explicit checks. The runtime separates work from acceptance, keeps ownership visible, constrains the admission verifier to a read-only role, resolves ambiguity fail-closed, and treats recovery as part of the control path. The research question is deliberately narrow: given the available evidence, what can this implementation support about auditable, verify-gated completion? We answer using event-level verify outcome summaries, PGV shadow summaries, and structural consistency checks.

The paper makes two connected contributions. First, it presents a five-plane admission-control design with explicit acceptance semantics, packetized state, read-only verification, fail-closed completion rules, and bounded recovery. Second, it reports a limited internal snapshot that supports event-level and structural claims about the control path.

The specific contributions are:

\begin{enumerate}
    \item We define a five-plane architecture that separates runtime work, authority and admission, context and memory, verification and recovery, and harness and learning concerns.
    \item We formalize acceptance semantics that distinguish execution, claim, acceptance, completion, blocked states, failure, recovery, and rollback, and map those states to packet fields and released event projections.
    \item We introduce a packetized state model with context shaping, memory ownership rules, and decision traces that support auditability.
    \item We add an evidence-admissibility framing that separates claims supported by the current supplementary accounting and schema package, case-study observations, and unsupported claims.
    \item We report a bounded empirical snapshot from one reference implementation together with an ancillary second-project procedure note to show which protocol elements transfer to a second setting, which remain scaffolded, and which claims stay out of scope.
\end{enumerate}

Sections~\ref{sec:problem} through~\ref{sec:readiness} develop these points from the formal model to the bounded empirical evidence.

The argument is not that control layers rescue weak models. The claim is narrower: once a system coordinates tools, roles, and long-lived state, admission and verification should be part of the runtime rather than an afterthought.

\noindent\textbf{Preprint scope note.} This preprint reports a bounded architecture case study. The released evidence supports inspectability and accounting claims about a verify-gated control path. It does not establish task-level success, comparative advantage over ungated baselines, production reliability, recovery effectiveness, or external validity.

The evidence should be read with its source limits in mind. The evidence base is internal and draws on an architecture handbook, a portable evidence digest, and associated runtime notes. The implementation observations reported in Sections~\ref{sec:evidence} and~\ref{sec:readiness} use an internal evidence-freeze package assembled from operator-facing snapshots, aggregate event summaries, runtime notes, and one ancillary second-project procedure note. Readers can inspect released verify, PGV, subset-accounting, and concentration summaries through the supplementary accounting and schema package. Event-taxonomy parity, configuration lint, and monitored consistency remain internal audit paths. The released accounting material provides anonymized schema material and compact accounting notes rather than a raw public event export. We therefore focus on checks tied directly to admission control: denominator hygiene, packet integrity, verification outcomes, and schema/config parity.

Before turning to related work, Table~\ref{tab:claims-scope} states the paper's claim boundaries. The rows map directly to the available evidence: released verify and concentration summaries for event-level verification, restricted audits and lint outputs for structural integrity, PGV shadow summaries for advisory governance signals, architecture and control-path material for recovery, and an ancillary second-project procedure note for procedural portability.

\begin{table}[!htbp]
\caption{Claims made and not made in this paper}\label{tab:claims-scope}%
\centering
\footnotesize
\renewcommand{\arraystretch}{1.12}
\begin{tabularx}{\linewidth}{L{0.18\linewidth}Y Y}
\toprule
Claim type & This paper reports & This paper does not claim \\
\midrule
Event-level verification & Verify success among invoked verify-completed events & End-to-end task success \\
Structural integrity & Event-taxonomy parity, configuration lint, and monitored consistency results & Correctness of every underlying decision \\
PGV & Shadow advisory signal & Runtime admission authority \\
Recovery & Architecture and control path & Proven recovery effectiveness \\
Generalization & Single-operator case-study evidence plus one ancillary second-project procedure note & Multi-project or production-wide validity \\
\bottomrule
\end{tabularx}
\end{table}

\section{Related work and the design gap}\label{sec:related}%

Recent agent work has expanded the orchestration toolkit. ReAct pairs reasoning with tool use in a single loop~\cite{react}. CAMEL and AutoGen emphasize role-specialized coordination~\cite{camel,autogen}, while MetaGPT extends that idea toward a role-structured software organization~\cite{metagpt}. SWE-agent shows the value of tighter coupling between agents and developer tools in software-engineering settings~\cite{sweagent}. These studies explain how work can be decomposed, coordinated, and grounded. They are less explicit about completion authority: who may declare the work done, what evidence must accompany that declaration, and what audit trail remains afterward.

A second line of work studies critique and revision. Reflexion reuses prior mistakes as verbal feedback~\cite{reflexion}. Self-Refine shows that self-feedback can improve outputs without additional training~\cite{selfrefine}. Chain-of-Verification structures checking prompts to reduce hallucinated or unsupported claims~\cite{cove}, and Constitutional AI examines rule-based critique as a way to shape behavior without relying only on outcome labels~\cite{constitutional}. This literature shows how checking can be inserted into an agent loop. The design studied here uses a narrower form of checking: the verifier sits outside the work step, is limited to admission, and does not rewrite the work product it evaluates.

Memory and observability become systems problems once work spans many turns, tools, and roles. MemGPT treats memory management that way rather than as a simple prompt-length issue~\cite{memgpt}. Dapper remains a useful reference for end-to-end tracing in distributed workflows, particularly when execution spans multiple components and no single local log is sufficient~\cite{dapper}. Agent systems face similar pressure when context moves across turns, tools, and specialized lanes.

Outside the agent literature, software engineering and operations have long treated release decisions as gated handoffs rather than as extensions of generation alone. Continuous-integration and continuous-delivery practice use acceptance tests and quality gates to decide whether a change may advance~\cite{humblecd,accelerate}. Organizational and IT-governance literature likewise treats decision rights and accountability splits as explicit coordination mechanisms rather than informal conventions~\cite{weillross}. Our architecture inherits that tradition: completion is not merely produced, but admitted through a bounded authority path with explicit evidence and an append-only audit trace.

Evaluation work also motivates careful denominators. AgentBench studies LLM agents in interactive environments, GAIA evaluates general assistants on real-world tool-using tasks, and judge-based work such as MT-Bench and Chatbot Arena explores scalable LLM evaluation while surfacing evaluator biases~\cite{agentbench,gaia,llmjudge}. More recent evidence shows that LLM evaluators can themselves be inconsistent and biased~\cite{llmevalbias}. This matters here because event-level pass rates, task-level coverage, and shadow-judge agreement are not interchangeable. We keep verify counts, task-level coverage, and shadow-evaluation outputs in separate evidence slices rather than forcing them into a single denominator.

Public frameworks now treat orchestration as runtime infrastructure. In the cited documentation snapshots, LangGraph emphasizes durable execution, human-in-the-loop support, and memory. CrewAI describes flows as structured, event-driven workflows that manage state and control execution together with collaborative crews. OpenAI's Agents SDK documents guardrails, human review, and run-control surfaces for when a run should continue, pause, or stop, while Semantic Kernel presents collaborative agent patterns for applications~\cite{langgraph,crewai,openaiguardrails,openaihitl,semantickernel}. These frameworks could host a verify-gated design, but the cited overviews foreground durable execution, flow control, guardrails, and collaboration more than completion authority itself.

Across these literatures, the handoff between acting and accepting is often underspecified. Many systems explain who performs the work more clearly than who decides that the work counts. Critics, judges, and reflection loops may exist, yet their authority can remain ambiguous. The design gap addressed here is not orchestration in general, but the treatment of packetized evidence, read-only admission authority, and denominator-aware completion accounting as first-class runtime concerns.

\section{System method and problem formulation}\label{sec:problem}%

Consider a task processed by a set of specialized lanes \(\mathcal{L}\), a packet store \(\mathcal{P}\), and an event ledger \(\mathcal{E}\). At time \(t\), the task state is represented as

\begin{equation}
\mathcal{T}_t = \langle g_t, \kappa_t, o_t, a_t, s_t, q_t, r_t, m_t \rangle,
\label{eq:taskstate}%
\end{equation}

where \(g_t\) is the current objective, \(\kappa_t\) is the accepted success criteria, \(o_t\) is the responsible owner, \(a_t\) is the accountable party, \(s_t\) is the current stage, \(q_t\) is the set of unresolved questions, \(r_t\) is the risk state, and \(m_t\) denotes the memory and packet references used to reconstruct the current task reality.

Table~\ref{tab:state-predicate-map} makes explicit how this conceptual task-state tuple is read by the later admission predicates. The tuple is not intended to enumerate every machine field in the implementation; it names the state dimensions that the packet and event fields must make inspectable.

\begin{table}[!htbp]
\caption{Mapping between task-state tuple components and admission predicates}\label{tab:state-predicate-map}%
\centering
\scriptsize
\setlength{\tabcolsep}{3pt}
\renewcommand{\arraystretch}{1.10}
\begin{tabularx}{\linewidth}{L{0.14\linewidth}L{0.32\linewidth}Y}
\toprule
Tuple component & Admission/control use & Runtime artifact or field family \\
\midrule
\(g_t,\kappa_t\) & Objective and success-criteria alignment for \(\phi_1\) & Common-ground packet, accepted facts, success criteria \\
\(o_t,a_t\) & Ownership/accountability and veto handling for \(\phi_5\) and \(\phi_{11}\) & Control header, owner/accountable fields, veto state \\
\(s_t\) & Stage and verify-mode state for \(\phi_2\) and \(\phi_3\) & Claim state, verify state, public \path{verify_completed} event \\
\(q_t\) & Open questions and unresolved blockers for \(\phi_7\), with escalation through \(\phi_8\) & Open questions, blocked reason, diagnostic-review fields \\
\(r_t\) & Risk, advisory, recovery, and rollback controls for \(\phi_8\)--\(\phi_{11}\) & Tier, advisory warning, recovery packet, rollback/veto state \\
\(m_t\) & Evidence, freshness, and packet lineage for \(\phi_4\) and \(\phi_6\) & Evidence packet, stale-ground/hash state, packet identifiers \\
\bottomrule
\end{tabularx}
\end{table}

Throughput is not the design target by itself. The target is completion that can be accepted under bounded risk, accountable ownership, and replayable evidence. This target leads to six objectives.

\begin{itemize}
    \item Separate execution from acceptance so that work done and work admitted are not conflated.
    \item Keep ownership explicit through responsible and accountable roles, with a distinct veto path for completion.
    \item Represent state through durable packets instead of relying on transient chat memory.
    \item Surface blocked and recovery states instead of forcing every run into a success or failure narrative.
    \item Keep advisory mechanisms advisory unless the protocol promotes them into authority.
    \item Limit governance overhead through tiering so that ceremony scales with task risk.
\end{itemize}

Several operational invariants follow. The runtime coordinator may route and synthesize, but it does not silently become the default worker. Every completion claim passes through a verify gate. Low confidence, weak evidence, missing ownership, skipped verification, unsupported states, and stale common ground all resolve fail-closed. Deep escalation follows a diagnostic-review-before-board path. Working context remains separate from research context, the admission verifier stays read-only, and recovery is a separate branch that re-enters the verify gate only through an explicit recovery packet and owner. Governance is retained only where it provides measurable value relative to its overhead.

\subsection{Bounded threat and failure model}

The architecture targets a bounded set of control failures rather than arbitrary model error. The focus is on failure patterns that can produce unsafe completion claims, hidden state drift, or loss of accountability in multi-agent orchestration. Table~\ref{tab:failure-model} summarizes the main failure modes, the intended control path, what the current evidence supports, and what remains unevaluated.

\begin{table}[!t]
\caption{Bounded failure model used in the case study}\label{tab:failure-model}%
\centering
\scriptsize
\setlength{\tabcolsep}{3pt}
\renewcommand{\arraystretch}{1.12}
\begin{tabularx}{\linewidth}{L{0.19\linewidth}L{0.20\linewidth}L{0.22\linewidth}Y}
\toprule
Failure mode & Primary control & Evidence currently available & Remaining evidence gap \\
\midrule
Premature completion claim & Verify gate, fail-closed admission, claim-accounting monitor & Event-level verify metrics, released case accounting outside the success headline, and fail-closed admission semantics & No aligned task-level coverage denominator or end-to-end task-success estimate \\
Weak or unsupported evidence & Evidence packet, diagnostic-review-before-board escalation, blocked state & Packet contract, blocked semantics, and conservative PGV shadow signaling are documented & No external adjudication study; the historical case table contains seven production blocked cases with generic \path{verify_blocked} reasons, so fine-grained root-cause attribution remains unavailable \\
Stale common ground or context drift & Common-ground snapshots, freshness checks, challenge/recovery path & Context separation and clean consistency scans support structural enforceability & No longitudinal drift-ablation study \\
Ambiguous ownership or advisory override & Ownership roles, veto path, admission verifier & Role taxonomy and acceptance rules are documented; PGV remains shadow-only & No released counts for ownership conflicts or advisory misses \\
Recovery or rollback path not exercised & Recovery packet, rollback queue, human-reviewed rollback policy & Recovery taxonomy and rollback bridge are documented & No real rollback events in the current corpus, so recovery effectiveness remains unevaluated \\
\bottomrule
\end{tabularx}
\end{table}

\section{System method: five-plane architecture}\label{sec:architecture}%

The proposed system is organized into five planes, summarized in Table~\ref{tab:planes}. The separation is operational, not merely diagrammatic. Agentic systems can accumulate hidden logic quickly: routing, memory, verification, rollback, and learning may blur into one another unless their responsibilities are named. The five-plane model keeps those boundaries visible. Throughout the paper, the shadow advisory completion-admission checker is referred to as the Policy/Governance Verifier (PGV).

\begin{figure}[!t]
\centering
\includegraphics[width=\linewidth]{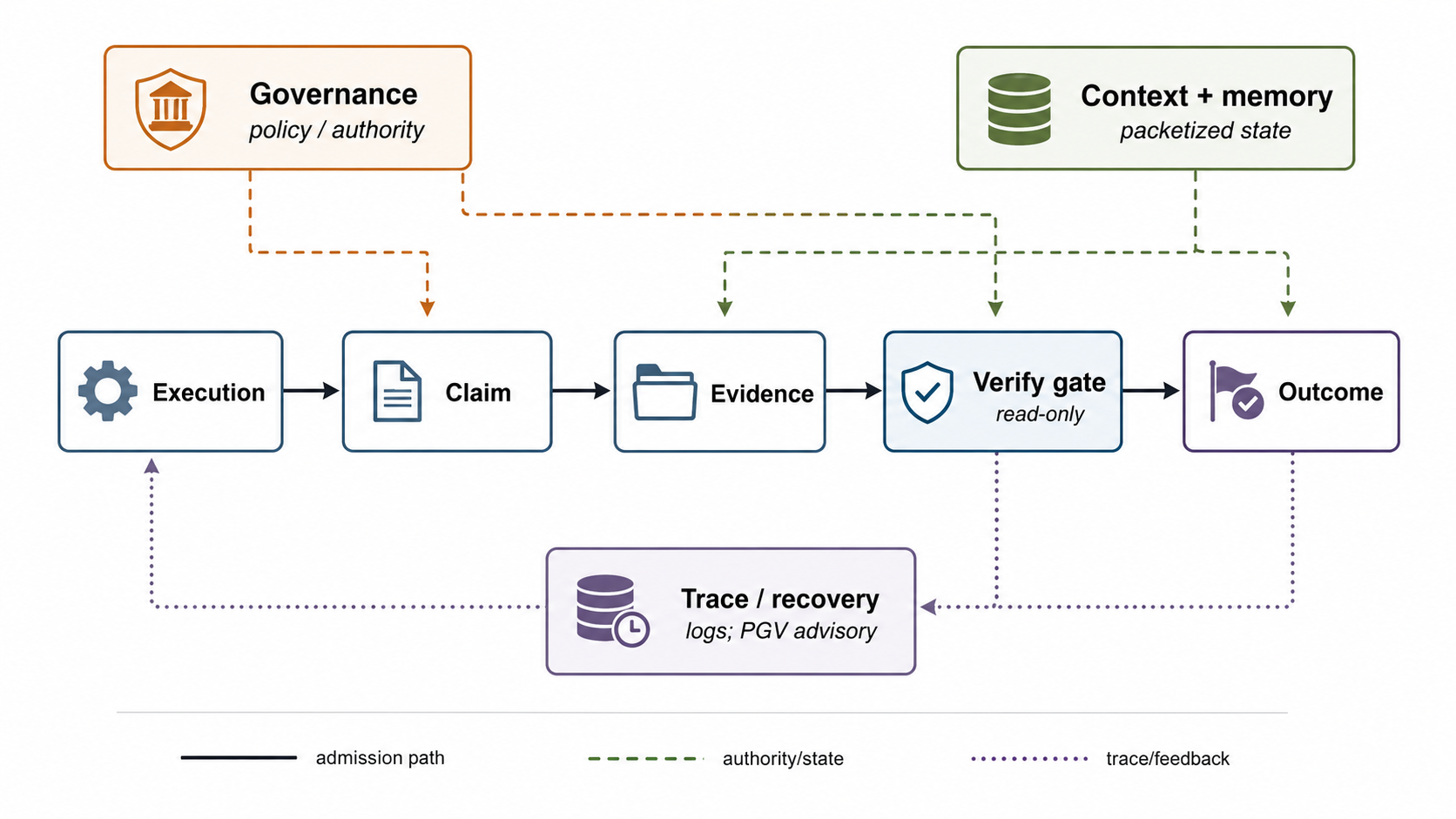}
\caption{Five-plane control surface. Solid arrows show the packet/admission path from execution through claim, evidence, verification, and outcome. Dashed arrows show authority and state constraints from the governance and context planes. Dotted arrows show trace, feedback, and advisory signals. Recovery connects back to the admission path when a blocked claim enters the recovery cycle, shown in detail in Fig.~\ref{fig:flow}.}\label{fig:planes}%
\end{figure}

\begin{table}[!t]
\caption{Five-plane architecture for governance-aware orchestration}\label{tab:planes}%
\centering
\footnotesize
\renewcommand{\arraystretch}{1.12}
\begin{tabularx}{\linewidth}{L{0.17\linewidth}L{0.20\linewidth}Y}
\toprule
Plane & Primary role & Representative abstract components \\
\midrule
Operational plane & Executes work and routes tasks & Runtime coordinator, implementation worker, discovery analyst, evidence retriever, design specialist, diagnostic reviewer, escalation board \\
Governance plane & Constrains work and decision rights & Common-ground policy, decision rights, premortem, challenge, dissent handling, trust calibration, after-action review \\
Context and memory plane & Maintains shared task reality and packets & Control header, common-ground packet, claim packet, evidence packet, recovery packet, context compiler, memory ownership rules \\
Verification and recovery plane & Decides whether a claim may surface and how recovery proceeds & Admission verifier, verify gate, deterministic evidence harnesses where available, exception classes, recovery policy, rollback policy, goal/progress monitor \\
Harness, measurement, and learning plane & Records execution, metrics, and bounded improvement & Event-first logging, governance trace, governance monitor, procedure packs, playbooks, PGV advisory, and operational snapshots \\
\bottomrule
\end{tabularx}
\end{table}

These labels denote abstract architectural roles. Here common-ground policy in the governance plane names the authority-bearing acceptance state, while the common-ground packet in the context and memory plane is the packetized artifact that records that state for routing and verification. The reference runtime instantiates these roles with implementation-specific lane names reported in Supplementary Information~1, but the architecture does not require those labels.

\subsection{Roles and decision rights}

Role specialization matters here, but the control point is that each role carries a boundary and a decision right. Table~\ref{tab:roles} lists the main abstract roles used in the paper.

\begin{table}[!t]
\caption{Core abstract roles and their control authority}\label{tab:roles}%
\centering
\scriptsize
\setlength{\tabcolsep}{3pt}
\renewcommand{\arraystretch}{1.15}
\begin{tabularx}{\linewidth}{L{0.16\linewidth}L{0.22\linewidth}Y}
\toprule
Abstract role & Main responsibility & Control authority and boundary \\
\midrule
Runtime coordinator & Ingests requests, selects tier, routes work, preserves continuity & Keeps owner and accountable fields current, opens the verify path, and surfaces outcomes, but does not silently absorb implementation work \\
Diagnostic reviewer & Handles deep diagnosis, ambiguity, and high-risk technical judgment & Can require premortem, challenge, or rerouting; deep escalation reaches the escalation board only through a diagnostic-first path \\
Implementation worker & Implements changes and runs focused execution loops & Can propose claims and evidence, but cannot self-admit completion \\
Discovery analyst & Maps repository structure and impact surface & Supports routing and scoping; does not hold release authority \\
Evidence retriever & Collects external or documentary evidence & Improves evidence quality and fact alignment; stays distinct from acceptance authority \\
Design specialist & Owns interface and interaction-level design decisions & Responsible for design artifacts, not final verification decisions \\
Admission verifier & Checks whether completion conditions are satisfied under a read-only contract & Holds the effective veto on completion claims; does not mutate the work product while verifying \\
Escalation board & Resolves ambiguity that remains after diagnostic review & Escalation board, not default path; used only after diagnostic-first processing \\
State integrity steward & Maintains versioning, staleness, archival integrity, and cleanup rules & Conceptual service role that protects state consistency without making business or technical task decisions \\
\bottomrule
\end{tabularx}
\end{table}

The decision model distinguishes responsible, accountable, consulted, informed, and veto relationships. When a claim is blocked, the next question is broader than what failed. The runtime must also identify who owns the next action, who may stop an unsafe outcome, and which roles remain advisory.

\subsection{Tiering and escalation}

The runtime operates in at least three tiers. Light mode handles narrow tasks with well-defined goals and little ambiguity. Standard mode supports multi-step synthesis and coordination, typically with up to two specialists. Deep mode is reserved for high-risk or ambiguous tasks. In deep mode, diagnostic-review-before-board becomes mandatory, premortem and challenge become available, and rollback policy is declared before execution gets too far.

The tier contract is intentionally fail-closed. Low-confidence results escalate to diagnostic-first review, as do unsupported or unclear requests. Blocked or skipped outcomes may not be surfaced as success. A next action without an explicit owner remains unresolved. Likewise, a worker report of \texttt{partial} or \texttt{not\_fixed} may not be converted into a completion claim by orchestration alone.

Table~\ref{tab:illustrative-trace} gives a compact synthetic and anonymized trace to make the admission path concrete. It is illustrative only; it is not an additional empirical sample.

\begin{table}[H]
\caption{Illustrative anonymized admission trace}\label{tab:illustrative-trace}%
\centering
\footnotesize
\renewcommand{\arraystretch}{1.12}
\begin{tabularx}{\linewidth}{L{0.32\linewidth}L{0.20\linewidth}Y}
\toprule
Step & Packet/Event & Admission meaning \\
\midrule
\path{task_created} & Common-ground packet & Objective, owner, and success criteria are initialized \\
\path{claim_packet_created} & Claim packet & Worker proposes completion, but acceptance remains pending \\
\path{verify_completed}: \texttt{blocked} & Verify outcome & Claim is withheld because a required predicate remains unresolved \\
\path{recovery_packet_created} & Recovery packet & Owner and next action are assigned for repair \\
\path{verify_completed}: \texttt{success} & Admission outcome & Completion is admitted after the repaired claim satisfies the gate \\
\bottomrule
\end{tabularx}
\end{table}

\section{System method: governance and acceptance semantics}\label{sec:semantics}%

The architecture distinguishes execution, claim, acceptance, and completion. Execution is the work itself. A claim is a structured assertion that the current state may satisfy the objective. Acceptance is the judgment that the claim meets the system's conditions for surfacing. Completion is the externally visible result of that judgment. This separation is central to the design.

\subsection{Acceptance rule}

For a claim \(c_t\) at time \(t\), the system accepts the claim only when the following predicate holds:

\begin{equation}
\acceptstate(c_t) = \mathbf{1}[\phi_1 \land \phi_2 \land \phi_3 \land \phi_4 \land \phi_5 \land \phi_6 \land \phi_7 \land \phi_8 \land \phi_9 \land \phi_{10} \land \phi_{11}],
\label{eq:accept}%
\end{equation}

where the conditions are: common ground is aligned with the current objective and success criteria (\(\phi_1\)); a claim packet exists and declares a completion-eligible done state rather than \path{partial} or \path{not\_fixed} (\(\phi_2\)); verification was invoked in the required mode (\(\phi_3\)); evidence meets the required floor (\(\phi_4\)); owner and accountable fields are clear (\(\phi_5\)); stale ground has been refreshed or ruled out (\(\phi_6\)); no blocked reason remains unresolved (\(\phi_7\)); any required deep-task escalation followed the diagnostic-review-before-board path (\(\phi_8\)); serious advisory warnings have been treated or explicitly dismissed under policy (\(\phi_9\)); no active recovery state remains (\(\phi_{10}\)); and no active veto condition is present (\(\phi_{11}\)).

The notation maps governance semantics to observable runtime artifacts; it is not intended as a theorem-bearing model. Operationally, \(\phi_1\) reads from the current common-ground packet; \(\phi_2\) from the claim packet; \(\phi_3\) from verify-mode state; \(\phi_4\) from the evidence packet; \(\phi_5\) through \(\phi_7\) from ownership, staleness, and blocker fields; and \(\phi_8\) through \(\phi_{11}\) from escalation, advisory, recovery, and veto control state. The public functions \(\omega_t(c_t)\) and \(\pi_t(c_t)\) correspond directly to the released \path{verify_completed}.\path{outcome} and \path{acceptance_status} fields.

The acceptance predicate determines both the public \path{verify_completed} outcome and the coarser public \path{acceptance_status} projection:

\begin{subequations}\label{eq:blockfail}
\begin{align}
\omega_t(c_t)
&=
\begin{cases}
\successout, & \acceptstate(c_t)=1, \\
\skippedout, & \neg \acceptstate(c_t) \land \sigma_t(c_t), \\
\blockedout, & \neg \acceptstate(c_t) \land \neg \sigma_t(c_t) \land \recoverable(c_t), \\
\failedout, & \neg \acceptstate(c_t) \land \neg \sigma_t(c_t) \land \neg \recoverable(c_t),
\end{cases}
\label{eq:blockedbranch}\\
\pi_t(c_t)
&=
\begin{cases}
\acceptedflag, & \omega_t(c_t)=\successout, \\
\withheldflag, & \omega_t(c_t) \in \{\blockedout,\failedout,\skippedout\},
\end{cases}
\label{eq:failedbranch}%
\end{align}
\end{subequations}

Here \(\omega_t(c_t)\) is the public \path{verify_completed} outcome and \(\pi_t(c_t)\) is the coarse public \path{acceptance_status} projection. The skip predicate \(\sigma_t(c_t)\) denotes a report-boundary non-release condition: verification ran and did not release completion, but downstream control flow has not yet committed the branch to blocked recovery or explicit stop. The predicates in Eq.~\ref{eq:blockfail} are evaluated in the displayed order: \(\sigma_t(c_t)\) is true only before the branch has been classified as recoverable or unrecoverable, so \(\neg\sigma_t(c_t)\land\recoverable(c_t)\) and \(\neg\sigma_t(c_t)\land\neg\recoverable(c_t)\) partition the non-skipped, non-accepted branches. Blocked therefore means ``not acceptable yet, but still recoverable in scope''. Failed means the current branch cannot continue safely within the present scope. Here \(\recoverable(c_t)\) is a predicate on the current claim branch rather than a lifecycle state. Rolled-back is different again: it is a safety state, not a success state. Operationally, Eq.~\ref{eq:accept} is fail-closed: until all mandatory predicates hold, the claim remains non-admitted and can proceed only through \path{skipped}, blocked recovery, failed stop, rollback, or later re-verification paths.

\subsection{Acceptance states}

Table~\ref{tab:states} summarizes the minimum internal branch states defined by the architecture. The public \path{verify_completed} outcomes and the coarse \path{acceptance_status} projection are separated in Table~\ref{tab:outcome-crosswalk} so that internal branch state is not conflated with public reporting markers.

\begin{table}[!t]
\caption{Minimum internal branch states in the architecture}\label{tab:states}%
\centering
\scriptsize
\setlength{\tabcolsep}{3pt}
\renewcommand{\arraystretch}{1.10}
\begin{tabularx}{\linewidth}{L{0.20\linewidth}L{0.24\linewidth}Y}
\toprule
Internal branch state & Meaning & Entry condition \\
\midrule
In-progress & Task is still being worked & \path{task_created}; no claim packet yet \\
Claim-ready & A structured claim exists & \path{claim_packet_created} \\
Verify-pending & Verification is running or awaiting evidence & \path{verify_started} \\
Verified-success & All mandatory admission predicates passed; completion is admitted for surfacing & \path{verify_completed}; outcome success \\
Blocked & Claim is not yet admissible but the task remains recoverable & \path{verify_completed}; outcome blocked, or \path{task_blocked} \\
Failed & Current branch is unsafe or infeasible in scope & \path{verify_completed}; outcome failed, or \path{task_failed} \\
Recovered & System went through recovery and returned to processing & \path{recovery_packet_created}; then \path{verify_started} resumes \\
Rolled-back & System returned to a known safe state & Rollback triggered and validated \\
\bottomrule
\end{tabularx}
\end{table}

The table is intentionally minimal and lists only internal branch states. In the supplementary accounting and schema package, \path{acceptance_status} is a coarse public projection rather than a separate lifecycle state. There, \path{accepted} is the public rendering of a successful \path{verify_completed} decision, not a second post-verification gate. By contrast, \path{withheld} covers blocked, failed, skipped, and other non-admitted claim states. It should therefore not be read as a synonym for \path{skipped}.

Table~\ref{tab:outcome-crosswalk} makes the layer separation explicit. It aligns internal branch reading, the public \path{verify_completed} outcome, and the coarser public flag \path{acceptance_status}. Because \path{success} is emitted only after the acceptance predicate passes, its mapping to \path{accepted} is deterministic. Because \path{withheld} is many-to-one, the same public flag appears on more than one distinct control path.

\begin{table}[!t]
\caption{Crosswalk from branch reading to public outcome and acceptance projection}\label{tab:outcome-crosswalk}%
\centering
\scriptsize
\setlength{\tabcolsep}{3pt}
\renewcommand{\arraystretch}{1.10}
\begin{tabularx}{\linewidth}{L{0.24\linewidth}L{0.20\linewidth}L{0.14\linewidth}Y}
\toprule
Branch reading & \shortstack[l]{\path{verify_completed}.\\\path{outcome}} & \shortstack[l]{\path{acceptance_}\\\path{status}} & Control-path consequence \\
\midrule
Verified-success & \path{success} & \path{accepted} & Claim is admitted and may surface as completion; \path{acceptance_status} simply projects that decision publicly \\
Blocked & \path{blocked} & \path{withheld} & Claim remains non-admitted; a recovery packet and a next recovery owner are required \\
Failed & \path{failed} & \path{withheld} & Current branch is non-admitted and unsafe in scope; the branch stops or enters supervised rollback review \\
Boundary non-release before downstream branch resolution & \path{skipped} & \path{withheld} & Verification did not release completion, so the claim stays non-admitted until later control flow commits the branch to rerouting, recovery, or explicit stop \\
\bottomrule
\end{tabularx}
\end{table}

Several fail-closed triggers are mandatory. A claim may not receive a \path{success} verification outcome or surface as completion if any of the following conditions hold:

\begin{itemize}
    \item verification was not invoked;
    \item the claim packet is missing or underspecified;
    \item evidence is weak or missing required references;
    \item ownership is inconsistent;
    \item stale common ground remains unresolved;
    \item deep-task escalation skipped the diagnostic-first path;
    \item serious advisory warnings remain untreated; or
    \item recovery is still in progress.
\end{itemize}

These operational triggers are not extra rules beyond Eq.~\ref{eq:accept}; they are the implementation-facing manifestations of \(\phi_2\), \(\phi_4\), \(\phi_5\), \(\phi_6\), \(\phi_8\), \(\phi_9\), and \(\phi_{10}\).

The canonical verification outcomes are \texttt{success}, \texttt{failed}, \texttt{blocked}, and \texttt{skipped}. Including \texttt{skipped} matters because it prevents non-released verification paths from disappearing into missing data.

\subsection{Lifecycle logic}

The nominal flow is easy to state even if it is careful in practice:

\begin{enumerate}
    \item task ingest and tier selection,
    \item lane routing and common-ground initialization,
    \item execution,
    \item claim assembly and evidence assembly,
    \item verify gate, and then
    \item either accepted completion, blocked recovery, failed/skipped non-admission, or rollback.
\end{enumerate}

What makes this more than a standard agent loop is that each stage is tied to packets and events. A worker does not simply say that the task is done; the system must create the artifacts that would allow another lane or a human operator to inspect that claim later.

\section{System method: packetized state and context management}\label{sec:packets}%

This architecture treats state as an explicit artifact rather than something that lives only in conversational residue. Table~\ref{tab:packets} lists the main packet types.

\begin{table}[!t]
\caption{Core packets in the proposed architecture}\label{tab:packets}%
\centering
\footnotesize
\setlength{\tabcolsep}{4pt}
\renewcommand{\arraystretch}{1.15}
\begin{tabularx}{\linewidth}{L{0.17\linewidth}L{0.34\linewidth}Y}
\toprule
Packet & Required contents & Primary consumers \\
\midrule
Control header & Tier, task class, owner, accountable, verify state, stale-ground flag, advisory signals & All lanes, especially the runtime coordinator and admission verifier \\
Common-ground packet & Objective, accepted facts, open questions, assumptions, current owner, current stage, success criteria & Runtime coordinator, diagnostic reviewer, implementation worker, admission verifier \\
Claim packet & Claimed done state, fix status, evidence snapshot, owner, accountable, unresolved questions, verify state & Admission verifier, runtime coordinator, PGV advisory \\
Evidence packet & Supporting references, missing evidence, evidence quality, source type, accepted facts & Diagnostic reviewer, admission verifier, PGV advisory \\
Recovery packet & Error class, failure signal, retry count, fallback used, next recovery action, recovery owner & Runtime coordinator, admission verifier, governance monitor \\
Procedure pack & Task archetype, required protocol, lane sequence, prompt families, required packets, rollback profile & Runtime coordinator, context compiler, learning layer \\
\bottomrule
\end{tabularx}
\end{table}

The packet rules are intentionally strict. Every packet has an owner, a version, and a lifecycle. Once a hash or staleness mismatch appears, old and new packets are not blended. A claim packet cannot be backfilled after a free-form answer as though it were clerical cleanup. An evidence packet must record what is missing as well as what is present. A recovery packet must identify who owns the recovery branch. A typical lineage is common-ground packet to claim packet to evidence packet to verify outcome. If the claim packet were backfilled after an answer was already surfaced, that lineage would be blurred and later audit would be weaker. These are small constraints, but they materially reduce ambiguity: borderline claims are pushed back into explicit packet, recovery, or non-admission paths instead of being smoothed over in prose.

\begin{figure}[!t]
\centering
\includegraphics[width=\linewidth,height=0.82\textheight,keepaspectratio]{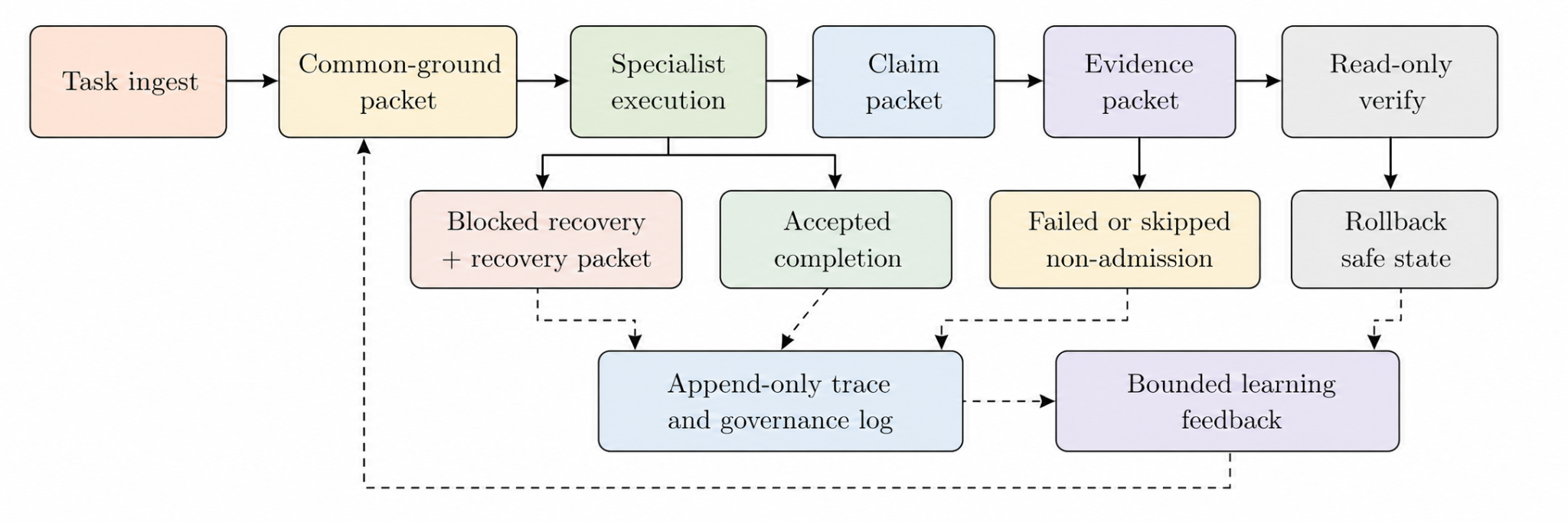}
\caption{Packet and decision flow for verify-gated completion. Solid arrows show the packet/admission path from task ingest through read-only verification to one admitted or non-admitted outcome; dashed arrows show event logging into the append-only trace; the dotted arrow shows reviewed learning feedback that stays off the hot path. Here \texttt{withheld} is the shared public flag for the blocked-recovery and failed/skipped branches.}\label{fig:flow}%
\end{figure}

\subsection{Working context, research context, and the context compiler}

The architecture also separates working context from research context. Working context contains the facts and state that may legitimately affect completion. Research context holds hypotheses, exploratory notes, and outside material that have not yet been promoted into accepted task reality. Keeping the two apart reduces the risk that tentative findings quietly become operational truth.

The context compiler shapes lane-specific context envelopes. It prefers packets to prose, narrow scope to long context dumps, and historical hints only when they are likely to reduce repeated failure or sharpen verification. In the reference design, an implementation worker receives the objective, exact scope, rollback hints, and recent verification misses when relevant. An admission verifier receives the claim packet, latest common-ground snapshot, stale and hash state, unresolved questions, and any minority note. A diagnostic reviewer receives accepted facts, assumptions, open questions, premortem input, and risk motifs. An evidence retriever receives the research question, the facts that need confirmation, and unresolved questions, but not the full working context by default.

\subsection{Memory ownership}

The memory model has three broad classes. Canonical memory includes common-ground snapshots, packets, after-action reviews, trust state, and governance labels that may act as control input. Archive-only memory includes raw logs, raw chat residue, synthetic or failed exploratory artifacts, and unreviewed notes; it is not injected directly into prompts. Prompt-injectable memory includes active governance hints, route corrections, verify hints, and procedure packs that have already been summarized and filtered.

This distinction matters because memory is both useful and risky. If everything becomes prompt-injectable, authority boundaries erode. If nothing persists beyond a task, the system cannot learn. The design therefore allows memory to influence routing and prompting, but only after ownership and memory class are explicit.

\section{System method: verification, recovery, and rollback}\label{sec:recovery}%

The verification and recovery plane is where the design differs most clearly from a typical agent loop. In many systems, critique and execution blur together. Here they are separated deliberately.

\subsection{Read-only verification}

The verify gate is read-only. It asks whether the claim packet, evidence packet, current common ground, and task state justify surfacing completion. It does not quietly rewrite the work product to make a borderline result pass. In the reference implementation, a \path{verify_completed} event with outcome \path{success} is emitted only when the full acceptance predicate in Eq.~\ref{eq:accept} holds; there is no second hidden post-verification admission stage, and the public \path{acceptance_status} field is only a coarse projection of that same decision. That boundary is reflected in role separation and event accounting: verification events inspect claim, evidence, and state packets, but they do not mutate the work product they evaluate. If the admission verifier silently repaired a failed patch and then emitted a \path{verify_completed} event with outcome \path{success}, the event trail would no longer distinguish checking from repair. That boundary matters because once the admission verifier becomes a worker, independent confirmation turns into self-repair.

In the reference implementation, the admission verifier remains the only admission authority, and the verify gate is the control point through which completion claims pass. The same shadow advisory completion-admission checker, referred to throughout as the Policy/Governance Verifier (PGV), runs as a rule layer that audits fail-closed preconditions without replacing the admission verifier at that gate. These PGV checks audit the same preconditions that must hold before a claim may emit \texttt{outcome = success}: populated ownership fields, present claim and evidence packets, no surfacing of partial or unfixed states as done, diagnostic-review handling for unclear requests, refresh of stale ground, and no active recovery state.

\subsection{Recovery as a first-class path}

Blocked states are treated as control outcomes, not side effects. When verification fails because evidence is weak, scope drift is unresolved, or ownership is unclear, the protocol routes the task back into recovery rather than treating it as nearly complete. The architecture requires a recovery packet, a recovery owner, and a named next action before the branch can return to verification. Only then can the task move toward accepted completion.

The goal and progress monitor follows the same principle. It does not replace orchestration. Its role is to surface early signs of objective drift, tier mismatch, stall, looped retries, or other patterns showing that the task is moving away from its success criteria. In the reference implementation this monitor remains partially scaffolded. That does not remove the need for the control path it is intended to support.

\subsection{Rollback}

The rollback path is intentionally bounded and human reviewed. The architecture identifies three representative triggers: high-risk verification failure, challenge-confirmed invalidity, and weak evidence discovered after a claim has been assembled. These triggers push advisory items into a rollback queue. An operator reviews the queue, declares rollback policy, and executes rollback if warranted. A successful rollback yields a validated safe state, while a failed rollback is logged as its own event. Rollback is therefore treated as a supervised operational action with its own trace.

\section{System method: traceability, metrics, and learning}\label{sec:trace}%

Traceability is necessary because governed agent systems are difficult to audit from local logs alone. Once multiple lanes, tools, and packets interact, the record must show how a completion claim reached admission. The architecture therefore uses event-first logging and a governance trace that ties runtime events back to the semantics of acceptance.

\subsection{Governance trace}

The minimum trace includes a run identifier, task identifier, event type, parent event identifier, stage transitions, owner and accountable fields, claim packet identifier, evidence packet identifier, verification outcome, acceptance status, blocked reason, protocol-expected and protocol-applied fields, latency where available, and event-count overhead proxies where true token or model cost is unavailable. In the reference implementation, packet identifiers for claim, evidence, recovery, and common-ground artifacts are threaded into later events so that verification and completion can be linked back to the exact packet lineage used at decision time. The trace does not yet provide true per-event token counts or model identifiers, so cost analysis remains pending instrumentation rather than a mature result claim.

This design borrows the idea of causal stitching from distributed tracing rather than adopting any one tracing tool. Dapper showed why causal stitching becomes necessary once work spans many services~\cite{dapper}. A governed agent runtime faces a similar problem. Without causal stitching, an operator sees a final answer and a few local logs; with it, the operator can reconstruct how a claim was formed, what evidence it cited, what verification concluded, and whether recovery or rollback changed the path.

\subsection{Measurement and accounting}

The architecture also defines a replay-accounting protocol. Session-based counting uses the session identifier in the event index as the authoritative scope for accountability. Rotation-aware accounting reconstructs the event sequence across both active and archived event files. For benchmark runs executed through the canonical reporting path, the ex post replay-accounting reconstruction identity is

\begin{equation}
\texttt{observed\_session\_events} = \texttt{expected\_events}.
\label{eq:benchmark}%
\end{equation}

Here \texttt{expected\_events} is reconstructed from the declared procedure pack for the canonical reporting path together with the realized branch profile observed in the event stream: it counts the packet-creation, verification, and outcome-reporting events required by that path, excludes operator-review artifacts outside the reporting path, and treats each recovery, rollback, or repeated verify cycle as an additional expected branch sequence rather than collapsing retries into one slot. This is therefore an ex post replay-accounting reconstruction identity rather than an independent integrity check or an ex ante compliance benchmark. A stricter ex ante minimal path can be derived from the declared procedure pack alone, but we do not report that quantity here. In this paper, we use the equality as a reporting convention rather than as an evaluated outcome metric. That distinction matters because it separates reconstruction fidelity from protocol compliance and keeps the reporting rules honest.

\subsection{Learning without runtime capture}

The learning loop runs outside the hot delivery path. Inputs include after-action reviews, governance violations, PGV shadow evaluations, and active governance memory. These inputs produce bounded candidate artifacts such as patch candidates, playbooks, active hints, and reasoning memory. Human review remains mandatory for promotion, finalization, retirement, and any movement from candidate status into reusable operating guidance.

That boundary is deliberate. Self-improvement is often presented as a natural extension of agent systems, but in operational settings it can become uncontrolled policy drift. This design allows learning only within a bounded scope. It does not imply autonomous code generation, automatic playbook application, or PGV-based runtime gating. Improvement is allowed, but the delivery path keeps its authority boundaries.

\section{Case-study methods}\label{sec:methods}%

We report a bounded case study centered on one primary reference implementation, plus an ancillary second-project procedure note used only as a portability check. The empirical aim is deliberately limited: show what the supplementary accounting and schema package can support, identify which claims remain internal-only, and keep each reported metric tied to the right denominator.

\subsection{Evidence sources and evidence slices}

The empirical material comes from four source groups: the internal architecture handbook, a portable evidence digest, operator-facing internal-check and consistency reports, and aggregate summaries derived from the event ledger and PGV shadow-evaluation artifacts. These materials are assembled into an internal evidence-freeze package rather than treated as a live operational feed.

Two evidence slices are used. The rotation-aware evidence slice supports the current event-level verification, production-split, and concentration summaries. A historical active-slice artifact set supports the internal-check corpus, the tier-overhead proxy, and the released nine-case failure table. We keep these slices separate because they answer different accounting questions and should not be merged into a single task-level denominator. Table~\ref{tab:slice-crosswalk} summarizes the slice boundaries used later in the Results section.

\begin{table}[!htbp]
\caption{Crosswalk between evidence slices used for event and case accounting}\label{tab:slice-crosswalk}%
\centering
\scriptsize
\setlength{\tabcolsep}{3pt}
\renewcommand{\arraystretch}{1.08}
\begin{tabularx}{\linewidth}{L{0.20\linewidth}L{0.20\linewidth}L{0.25\linewidth}Y}
\toprule
Slice or view & Used for & Counts referenced in this manuscript & Interpretation boundary \\
\midrule
Rotation-aware verify-completed evidence slice & Current event-level verify accounting, production split, and concentration & 1,801 verify-completed rows; 1,800 known outcomes; 8 production blocked rows; 1 synthetic rollback-drill failure; 1 synthetic missing-outcome artifact & Latest event-level denominator; supports aggregate accounting but not fine-grained root-cause attribution \\
Historical active-slice artifact set & Historical case-level table and tier-overhead proxy & Nine records outside the historical success headline: 7 production blocked cases, 1 synthetic rollback-drill failure, and 1 synthetic missing-outcome artifact; 785 verify-completed rows in approximately 8,240 standard-tier events & Historical artifact denominator; should not be merged with the rotation-aware event slice \\
Production-subset task view & Descriptive task-id subset accounting & 69 production task identifiers; 11 with at least one verify-completed event; 4 \path{task_completed} tasks, all 4 with verify-completed events & Bounded subset count only; not a global task-level coverage rate \\
\bottomrule
\end{tabularx}
\end{table}

We therefore report aggregate figures from the distilled package rather than releasing a raw public event export. To improve accounting transparency without exposing internal traces, Supplementary Information~1 (ESM\_1.pdf) and the supplementary accounting and schema package provide an anonymized event-schema sketch, a tabulated synthetic event sample, minimal verify-gate pseudocode, packet templates, denominator-provenance and subset-accounting notes, concentration and overhead notes, an event-taxonomy-parity note, and aggregate verify and PGV breakdowns. This distilled package preserves the released calculations and denominator logic, while avoiding operator-specific traces that would otherwise expose internal workflow details.

\subsection{Inclusion, exclusion, and units of analysis}

Three units of analysis are kept separate. Structural checks evaluate whether schemas, runtime events, and configuration surfaces remain synchronized. Event-level metrics quantify properties of individual verification events in the released verification slice. Implementation-status summaries aggregate operator-facing snapshots such as task counts, consistency scans, and the paper-facing internal-check corpus.

That internal-check corpus contains 88 checks, of which 72 are synthetic and 16 are non-synthetic. It is distinct from the rotation-aware evidence slice used for current event-level reporting, which contains 1,801 released verify-completed rows. In the compact accounting and schema package, this same paper-facing historical readiness material is occasionally labeled \path{completion_readiness_check}; throughout the manuscript, we refer to that bounded subset as the internal-check corpus. Because the available event corpus and task-summary corpus are not aligned by a common eligibility denominator, global task-level verify coverage is not computed in this draft.

\subsection{Metric definitions and Policy/Governance Verifier (PGV) protocol}

All percentages in the Results section are reported with explicit denominators. Event-level verify success is defined as successful verify-completed rows divided by all such rows with populated outcomes. In the rotation-aware evidence slice, that denominator is 1,800 after excluding one synthetic missing-outcome artifact from the all-row total of 1,801, because an event with no recorded outcome cannot contribute to outcome-conditioned success accounting. A single task may also trigger more than one \path{verify_completed} event; for example, a blocked verify-completed event after an initial claim can be followed by a successful verify-completed event after recovery. Event-level verify success should therefore not be read as a task-level completion rate.

We do not report global task-level verify coverage because the aligned corpora do not provide a denominator of completion-eligible tasks. Computing that denominator would require a stable join between task identifiers in the event ledger and the task-summary corpus, together with an explicit eligibility rule for completion-eligible tasks. The released accounting and schema package does not expose that join. Instead, we report the production subset separately: 69 production task identifiers were observed, 11 had at least one verify-completed event, and all four production tasks with \path{task_completed} also had a verify-completed event. These are descriptive subset counts rather than global coverage rates.

Structural integrity claims rely on three visibility levels. Event-taxonomy parity is available through a restricted internal audit path. Configuration lint is a point-in-time internal report. The monitored internal consistency checks come from operator-facing scan outputs rather than from independently replayable supplementary-package artifacts.

PGV, the same shadow advisory completion-admission checker described in Section~\ref{sec:recovery}, is analyzed only in shadow mode. The supplementary accounting and schema package supports aggregate shadow-evaluation outputs---rule agreement, false-success, blocked precision, and limited blocked recall notes---over 2,044 samples. The PGV denominator map used below is \(N_{\mathrm{all}}=2{,}044\), \(N_{\mathrm{final}}=1{,}548\) for comparable finalized-outcome rows, \(N_{\mathrm{safe\ pred}}=1{,}526\) for safe-to-proceed predictions, and \(N_{\mathrm{blocked\ pred}}=518\) for blocked/risky predictions. The two prediction-side counts partition the 2,044-sample pool; the 1,548-row finalized-outcome subset is used for rule-agreement accounting. The repeated value 1,526 therefore appears both as an agreement count and as a prediction-side subset size, and those roles are not interchangeable. Because the supplementary accounting and schema package alone does not provide a stronger external ground-truth procedure, we read the PGV results conservatively as internal shadow signals. That caution matters again in the Results section: high agreement alone is not enough to justify runtime authority, and the current blocked signal is not decision-useful in the evaluated setup. The evaluated subset is class-imbalanced, and the blocked precision of 2/518 = 0.39\% shows substantial over-flagging. Supplementary Information~1 (ESM\_1.pdf) reports both the separate PGV denominators and a denominator map spanning the 2,044-sample pool, the 1,548 comparable finalized-outcome subset, and the 1,526 safe-to-proceed / 518 blocked-risk prediction subsets so that class-imbalance risk remains visible. More informative blocked-performance estimates would require targeted sampling of blocked production cases together with blinded adjudication.

\section{Results: evidence framing and admissible claims}\label{sec:evidence}%

The case study draws on several kinds of material: handbook prose, operator checks, consistency scans, event-ledger analyses, and shadow evaluations. They do not all carry the same evidentiary weight. We therefore sort them into three classes. First are claims supported directly by the released supplementary materials. Second are bounded internal observations that remain visibility-limited and are not independently checkable from the supplementary accounting and schema package. Third are case-study-only or unsupported claims that should not be read as validated results. Denominator notes and the bounded failure-mode/control map are provided in Supplementary Information~1 (ESM\_1.pdf). The tables below keep those classes separate so that released-material support is not conflated with restricted audits, internal lint outputs, monitored windows, or historical artifacts.

\begin{table}[!htbp]
\caption{Claims supported by supplementary accounting materials}\label{tab:evidence-released}%
\centering
\scriptsize
\setlength{\tabcolsep}{3pt}
\renewcommand{\arraystretch}{1.02}
\begin{tabularx}{\linewidth}{L{0.19\linewidth}L{0.24\linewidth}L{0.12\linewidth}Y}
\toprule
Claim family & Concrete evidence in the supplementary accounting and schema package & Status & Allowed interpretation \\
\midrule
Known-outcome invoked-event verify success share & 1791/1800 known-outcome verify-completed rows = 99.5\% & Supported & Checkable event-level accounting over verify outcomes with populated outcomes; not a task-level or end-to-end success claim \\
All-row invoked-event verify success share & 1791/1801 all verify-completed rows = 99.44\% & Supported with caveat & Checkable all-row accounting including one synthetic missing-outcome artifact; denominator context only, not the headline accounting share \\
PGV shadow advisory evaluation & n=2044 total; 1,526/1,548 = 98.58\% rule agreement on the comparable finalized-outcome subset; 0/1,526 = 0.0\% false-success among safe-to-proceed predictions; 2/518 = 0.39\% blocked precision among blocked/risky predictions & Supported with caveats & Checkable shadow signal with a safe-pass bias, but the current blocked predictions are not decision-useful enough for runtime gating \\
Synthetic/session dominance & 1784/1801 verify-completed rows are synthetic, session, or otherwise non-production & Supported with framing & Checkable composition evidence showing the released verify slice is dominated by non-production events \\
Snapshot concentration & 1762/1801 verify-completed rows came from a single high-volume reporting cluster & Supported with framing & Checkable concentration evidence showing the released verify slice is snapshot-concentrated rather than longitudinally sampled \\
\bottomrule
\end{tabularx}
\end{table}

\begin{table}[!htbp]
\caption{Bounded internal observations reported separately from released-material findings}\label{tab:evidence-internal}%
\centering
\scriptsize
\setlength{\tabcolsep}{3pt}
\renewcommand{\arraystretch}{1.04}
\begin{tabularx}{\linewidth}{L{0.18\linewidth}L{0.20\linewidth}L{0.20\linewidth}Y}
\toprule
Observation & Slice / visibility & Concrete evidence & Allowed interpretation \\
\midrule
Fail-closed governance (monitored window) & Internal monitored evidence window / operator-facing monitors & 0 monitored violations, 0 monitored abandoned tasks, 0 monitored stalled tasks, 0 monitored rollback-pending tasks & No monitored governance violations were observed in the covered window; because the monitor surface is internal and not always-on across every control path, this remains a bounded internal observation rather than a system-wide fail-closed guarantee \\
Event taxonomy parity & Active-schema snapshot / restricted internal audit & Parity across active schema surfaces; the supplementary accounting and schema package exposes only a paper-facing schema sketch and parity note & Internal taxonomy-synchronization claim at snapshot time; not independently checkable from the supplementary accounting and schema package alone \\
Config parity & Point-in-time internal config state / internal lint report & Internal verify and main configuration check returned 0 failures & Point-in-time verify and main configuration integrity from internal lint output; not independently replayable from the supplementary accounting and schema package alone \\
Consistency scanner & Current monitored scan window / internal operator-facing scan & All monitored consistency checks reported 0 over 2255 tasks & Clean structural consistency in the monitored window; not proof that every underlying decision was correct \\
\bottomrule
\end{tabularx}
\end{table}

\begin{landscape}
\begin{table}[p]
\caption{Evidence grading for case-study-only and unsupported claims}\label{tab:evidence-limited}%
\centering
\scriptsize
\setlength{\tabcolsep}{2pt}
\renewcommand{\arraystretch}{0.98}
\begin{tabularx}{\linewidth}{L{0.15\linewidth}L{0.18\linewidth}L{0.22\linewidth}C{0.10\linewidth}Y}
\toprule
Claim family & Slice / visibility & Concrete evidence & Grade & Allowed interpretation \\
\midrule
Task-level verify coverage & Cross-corpus denominator unavailable / no released aligned join & Not currently computable from aligned project-attributed data; the earlier cross-corpus ratio was invalid & Unsupported & Coverage claims require an aligned numerator and denominator; current sources support event-level results only \\
Production-subset verify path & Observed production subset / supplementary subset-accounting summary & 11 observed production task IDs had at least one verify-completed event out of 69 observed production task IDs; all 4 production tasks with a \path{task_completed} event also had a verify-completed event & Case-study & Subset accounting only; not a global task-level coverage rate \\
Failure-mode interception (latest slice) & Latest rotation-aware evidence slice / released aggregate summary & 8 production blocked rows, 1 synthetic rollback-drill failure, and 1 synthetic missing-outcome artifact & Case-study & Supports fail-closed accounting in the latest released slice, but not fine-grained causal attribution \\
Failure-mode interception (historical cases) & Historical active-slice artifact set / released case-level table & Nine-case historical artifact & Case-study & Preserves case-level blocked/failed and missing-outcome examples for audit, but reflects a historical slice rather than the latest rotation-aware one \\
Overhead proxy by tier & Historical active-slice artifact set / released overhead note & Standard-tier events dominate; 785 verify-completed rows appear in roughly 8,240 standard-tier events, with one additional unknown-tier synthetic missing-outcome artifact & Case-study & Event-count proxy only; not latency or cost overhead \\
Generalization, comparative advantage, production reliability, learning effectiveness, recovery effectiveness & No multi-project cohort or rollback corpus / released and internal materials remain insufficient & No multi-project cohort, no ungated baseline, limited production verify-completed evidence, no usage telemetry, and no real rollback corpus & Unsupported & These belong in limitations or future work, not in the paper's result claims \\
\bottomrule
\end{tabularx}
\end{table}
\end{landscape}

This grading keeps the evidence boundaries visible. Table~\ref{tab:evidence-released} marks the findings that a reader can trace directly through the supplementary accounting materials. Table~\ref{tab:evidence-internal} separates bounded internal observations that remain visibility-limited. Table~\ref{tab:evidence-limited} marks what remains case-study only or unsupported. The clearest claims supported by released supplementary materials are event-level verify behavior among invoked runs, PGV shadow metrics, synthetic dominance, and snapshot concentration. Internal parity, lint, and monitored consistency checks remain contextual bounded observations rather than independently checkable results. The most consequential unsupported claim is any global task-level or production-reliability estimate, because the eligible denominator is not aligned. The same logic rules out stronger claims about external validity, comparative superiority, or proven learning effectiveness.

\subsection{Denominator policy and event-level verification}

All percentages in this section use explicit denominators. The main observable is event-level verify success among verify-completed rows with populated outcomes. Task-level coverage would require an aligned denominator of completion-eligible tasks, so we report bounded subset counts rather than a global coverage rate.

\begin{table}[!htbp]
\caption{Verification event accounting in the rotation-aware evidence slice}\label{tab:verify-denominators}%
\centering
\scriptsize
\setlength{\tabcolsep}{2pt}
\renewcommand{\arraystretch}{1.12}
\begin{tabularx}{\linewidth}{L{0.16\linewidth}C{0.08\linewidth}C{0.08\linewidth}C{0.08\linewidth}C{0.08\linewidth}C{0.08\linewidth}Y}
\toprule
Scope & Rows & Success & Blocked & Failed & Missing & Interpretation \\
\midrule
All rows & 1801 & 1791 & 8 & 1 & 1 & Includes one synthetic missing-outcome artifact in the rotation-aware evidence slice \\
Known-outcome rows & 1800 & 1791 & 8 & 1 & 0 & Canonical denominator for the known-outcome invoked-event verify success share \\
Production only & 17 & 9 & 8 & 0 & 0 & Small non-synthetic subset used only for bounded subset observations \\
Synthetic / session & 1784 & 1782 & 0 & 1 & 1 & Dominant benchmark, drill, test, or session-artifact slice in the rotation-aware evidence slice \\
\bottomrule
\end{tabularx}
\end{table}

The rotation-aware evidence slice contains 1,801 verify-completed rows. Of these, 1,800 have populated outcomes: 1,791 success, eight blocked, and one failed. One additional synthetic test artifact lacks an outcome field. We exclude that artifact from the canonical denominator because an event without a recorded outcome cannot support outcome-conditioned success accounting, but we retain it in the all-row note. Because the missing-outcome artifact is synthetic rather than production-classified, this filtering step does not remove a known production outcome from the reported rate. The known-outcome invoked-event verify success share is therefore 1791/1800 = 99.5\% among known-outcome verify-completed rows. If the missing-outcome artifact is included in the denominator, the all-row share is 1791/1801 = 99.44\%. This is an accounting measure over invoked verification events, not a task-completion, production-reliability, or benchmark-success rate. For example, one task can contribute a blocked verify-completed event before recovery and a later successful verify-completed event after the claim is repaired.

\begin{table}[!htbp]
\caption{Production-subset task accounting}\label{tab:production-subset}%
\centering
\footnotesize
\renewcommand{\arraystretch}{1.12}
\begin{tabularx}{\linewidth}{L{0.42\linewidth}C{0.14\linewidth}Y}
\toprule
Metric & Value & Claim status \\
\midrule
Unique task IDs in historical active event corpus & 1,234 & Corpus descriptor only; not an eligible task-level denominator \\
Production task IDs & 69 & Non-synthetic subset descriptor \\
Production tasks with at least one verify-completed event & 11/69 = 15.9\% & Descriptive subset share only; not a validated global coverage numerator \\
Production tasks with \path{task_completed} & 4 & Completed production subset \\
Completed production tasks with a verify-completed event & 4/4 & Narrow supported claim \\
Global task-level verify coverage & Not computable & Eligible task denominator is not aligned in the current corpora \\
\bottomrule
\end{tabularx}
\end{table}

Global task-level verify coverage is not computable from the current aligned corpora. A production-subset analysis identified 69 non-synthetic task identifiers, 11 of which (15.9\%) had at least one verify-completed event recorded. This descriptive subset accounting is not a global coverage estimate because the denominator includes tasks that may not have been completion-eligible. The narrower supported finding is that all four production tasks with \path{task_completed} also had a corresponding verify-completed event.

\subsection{Synthetic/non-production split and snapshot concentration}

The verify-completed slice (event type \path{verify_completed}) is synthetic-heavy. In the rotation-aware evidence slice, 1,784 of 1,801 such rows are synthetic, session, or otherwise non-production, while only 17 are production-classified. The production subset contains nine successes and eight blocked outcomes. The synthetic/session subset contains 1,782 successes, one failed synthetic drill, and one missing-outcome synthetic artifact. These data support structural and procedural claims, but they do not support broad production reliability claims.

The event-level verification evidence is also snapshot-concentrated. In that same slice, 1,762 of 1,801 verify-completed rows came from a single high-volume reporting cluster, leaving only 39 verify-completed rows outside that cluster. Here ``high-volume reporting cluster'' means one bounded burst of reporting activity grouped under a shared reporting context in the released slice, not a separate deployment tier or customer cohort. The pattern therefore suggests a batch-concentrated reporting slice rather than a longitudinally sampled operating period. We avoid longitudinal extrapolation beyond the observed slice.

\subsection{PGV shadow advisory evaluation}

PGV---the same shadow advisory completion-admission checker described earlier---was evaluated only as a shadow advisory signal. Supplementary Information~1 separates the total 2,044-sample evaluation pool from the 1,548 comparable finalized-outcome subset and the prediction-side denominators used for false-success (1,526 safe-to-proceed predictions) and blocked precision (518 blocked/risky predictions). Within that released aggregate view, PGV achieved 1,526/1,548 = 98.58\% rule agreement and 0/1,526 = 0.0\% false-success. However, blocked precision was 2/518 = 0.39\%, indicating that the current blocked signal is not decision-useful in the evaluated setup and over-flags heavily. The apparent blocked recall of 2/2 is limited by the very small number of actual blocked cases and is not treated as a full-corpus recall estimate. PGV is therefore not promoted to admission authority and is not reported as a usable blocking function.

\subsection{Blocked/failed and missing-outcome case accounting}

The released case-level accounting table classifies nine historical active-slice verify-completed records outside the historical success headline. One is a synthetic missing-outcome artifact, one is an expected synthetic rollback-drill failure, and seven are production gap-analysis tasks blocked with generic \path{verify_blocked} reasons. Because these generic \path{verify_blocked} reasons do not preserve class-specific root causes, the released records support fail-closed accounting but not fine-grained failure-mode attribution.

\subsection{Overhead proxies by tier}

Because wall-clock time, token counts, and model identifiers are not consistently instrumented, we do not report true latency or cost overhead. Instead, we report event-count proxies by tier from the historical active-slice artifact set. In that artifact set, standard-tier events dominate, and 785 verify\mbox{-}completed rows appear within approximately 8,240 standard-tier events (about 9.5\%), with one additional unknown-tier synthetic missing-outcome artifact recorded separately from the named tiers. These proxy metrics describe logging and event composition, not runtime latency or financial cost.

\subsection{Remaining evidence gaps}

Several gaps remain. The available corpus contains no real rollback events, so recovery effectiveness is still unevaluated. Production-classified verify-completed evidence is limited to 17 verify-completed events in the rotation-aware evidence slice.

Across that evidence slice, there are nine outcome-bearing non-success verify-completed rows overall (one synthetic rollback-drill failure and eight production gap-analysis rows blocked with generic \path{verify_blocked} reasons), plus one synthetic missing-outcome artifact. The compact released accounting and schema package preserves case-level accounting only for the historical nine-case active-slice artifact set, so these materials do not support fine-grained root-cause attribution.

True per-event token counts, model identifiers, and representative timing fields are not yet instrumented, so latency and cost overhead remain non-computable. The reasoning-memory pipeline contains 10 active memories, all at medium confidence, and the current package does not show whether those memories improved later decisions. Finally, no degraded-mode ablation comparing verify-gated admission against an ungated baseline was executed, so comparative effectiveness and relative risk reduction cannot be estimated.

\section{Implementation snapshot and portability note}\label{sec:readiness}%

The architecture handbook and evidence digest provide a bounded, operator-facing snapshot of the reference implementation. Table~\ref{tab:snapshot} retains only the measures most closely tied to the paper's research question: verification accounting, slice composition, PGV shadow behavior, and structural-alignment checks at snapshot time.

\begin{table}[H]
\caption{Core snapshot measures relevant to the research question}\label{tab:snapshot}%
\centering
\scriptsize
\setlength{\tabcolsep}{4pt}
\renewcommand{\arraystretch}{1.10}
\begin{tabularx}{\linewidth}{L{0.29\linewidth}Y}
\toprule
Metric & Observation \\
\midrule
Verify accounting & 1,801 all rows; 1,800 known-outcome rows \\
Known-outcome invoked-event verify success share & 1791/1800 = 99.5\% among known-outcome rows; 1791/1801 = 99.44\% across all rows \\
Production-classified verify-completed events & 17 production verify-completed events: 9 success and 8 blocked \\
Completed production subset & 4/4 production tasks with \path{task_completed} also had a verify-completed event; 11/69 observed production task IDs had at least one verify-completed event \\
Synthetic/session dominance & 1784/1801 released verify-completed rows are synthetic, session, or otherwise non-production \\
Snapshot concentration & 1762/1801 released verify-completed rows came from a single high-volume reporting cluster \\
PGV shadow advisory evaluation & n=2044 total; 1,526/1,548 = 98.58\% rule agreement on the comparable finalized-outcome subset; 0/1,526 = 0.0\% false-success among safe-to-proceed predictions; 2/518 = 0.39\% blocked precision among blocked/risky predictions \\
Structural integrity snapshot & Restricted-audit event-taxonomy parity across active schema surfaces; point-in-time config lint reported 0 failures; current monitored consistency scan reported 0 findings over 2255 tasks \\
\bottomrule
\end{tabularx}
\end{table}

These rows are best read as a compact operational snapshot, not as a maturity scorecard. The research-question-relevant measures remain event-level verify accounting, the size and composition of the released slice, PGV shadow behavior, and structural-alignment checks at snapshot time. They support inspectability and governance-alignment claims, not production-performance claims.

The main caution is compositional rather than procedural: the verify-completed slice is synthetic-heavy and concentrated, while the production subset remains small. PGV likewise stays advisory because high rule agreement coexists with strong blocked over-flagging. Although the runtime includes operational controls, these numbers remain bounded evidence rather than broad deployment validation.

Table~\ref{tab:maturity} summarizes the capability states most relevant to the admission-control path.

\begin{table}[H]
\caption{Selected capability maturity states most relevant to the control path}\label{tab:maturity}%
\centering
\footnotesize
\setlength{\tabcolsep}{4pt}
\renewcommand{\arraystretch}{1.15}
\begin{tabularx}{\linewidth}{L{0.24\linewidth}L{0.12\linewidth}Y}
\toprule
Capability & Status & Operational reading \\
\midrule
Context compiler & Side-path helper & Emits lane-shaped artifacts and is used when invoked, but does not sit on every main runtime path \\
Memory ownership charter & Live & Explicit memory events and ownership taxonomy are present in the runtime pack \\
Goal and progress monitor & Scaffold & Event types and partial specification exist, but no dedicated always-on monitor loop is reported \\
Hard rollback primitive & Scaffold to live & Detection and execution bridge exist with human review, but rollback is not automatically triggered \\
Completion-admission check & Live and advisory & Formalizes fail-closed rules as an advisory check and does not replace the admission verifier at the verify gate \\
Procedure packs and playbooks & Scaffold to live & Lifecycle commands exist, promotion is human reviewed, and runtime auto-selection remains intentionally bounded \\
\bottomrule
\end{tabularx}
\end{table}

The snapshot shows uneven but interpretable maturity. The controls around acceptance, traceability, and memory provenance are live or close to live, while monitoring and some recovery automation remain scaffolded. For a reliability-oriented system, this ordering is plausible: admission controls are established before additional automation is trusted.

\subsection{Ancillary second-project procedure note}

Apart from the primary reference corpus, the same verify-gated protocol was exercised on an anonymized second Rust workspace as a bounded portability check. Four tasks were executed: build verification, invariant audit, CI/Makefile parity scan, and lineage coverage scan. Workspace-wide build, lint, no-run test, and formatting checks passed without source modification. The audit also documented one future-proofing gap together with lineage and CI/Makefile parity gaps.

We treat this as an ancillary procedure note rather than as a core empirical result. It shows that the protocol can be applied coherently outside the primary corpus, but it does not measure transfer effectiveness, outcome quality, or multi-project generalization. Project identifiers remain anonymized because the note is used only to assess procedural portability, not to make project-specific claims.

\section{Discussion and limitations}\label{sec:discussion}%

This architecture fits best where an incorrect completion decision is costly, such as software changes, evidence-based research synthesis, regulated review workflows, and other long-lived tasks. In those settings, explicit admission rules and traceability can justify the extra structure.

Four design patterns summarize the architecture in practical terms. Admission Gate keeps completion authority separate from execution. Packetized Evidence Chain ties completion claims to named packets rather than free-form assertions. Fail-Closed Non-Admission makes blocked and failed branches explicit instead of smoothing them into soft success. Advisory Learning Without Authority keeps PGV and learning signals useful without letting them silently take runtime authority.

These patterns are not meant to replace software-engineering quality gates or deterministic tests. They adapt quality-gate and audit-trail discipline to agentic workflows, where completion claims are easy to blur if authority, evidence, and traceability are left implicit.

The architecture is a weaker fit for disposable or low-risk tasks. When the cost of error is low, the protocol overhead can outweigh the benefit. Packet assembly takes time, verification adds overhead, ownership has to remain explicit, and the event log needs upkeep. The design remains practical only if tiering is meaningful and the light tier stays light.

The current case study also says little about scale. It does not show what happens when hundreds of agents are active at once, when many claims are in flight together, or when coordination spans many projects. Governance overhead will likely grow with packet versioning, verification fan-out, and recovery-branch management. That does not refute the design, but it leaves scalability as an open systems question.

The architecture also has limits. It cannot turn weak models into strong ones, replace deterministic checks where those checks are available, rule out false negatives or overblocking, or compensate for poor task decomposition.

\noindent\textbf{Construct validity.} No degraded-mode ablation comparing verify-gated admission with an ungated baseline was executed, so comparative effectiveness and relative risk reduction remain out of scope. The reported verify rates are event-level outcomes among verify-completed rows with populated outcomes, not task-level coverage or end-to-end performance measures.

\noindent\textbf{Data validity.} The released slice is synthetic-heavy and concentrated: 1,784 of 1,801 verify-completed rows are non-production, and 1,762 of 1,801 come from one high-volume reporting cluster. This is best read as a concentrated snapshot rather than as a longitudinal sample of production behavior.

\noindent\textbf{External validity.} Production-classified verify evidence remains limited to 17 verify-completed events. The ancillary second-project procedure note shows portability of the procedure, not multi-project effectiveness.

\noindent\textbf{Measurement validity.} PGV remains advisory: blocked precision is 2/518 = 0.39\%, corpus-level recall and F1 are not computable from the supplementary accounting and schema package, and true latency, token, and model-cost instrumentation remains incomplete.

\section{Conclusion}\label{sec:conclusion}%

This manuscript is best read as an architecture and evidence-framing preprint, not as a benchmark, production-validation, or comparative performance study. In multi-agent systems, completion control depends on what the model can do, who is allowed to declare completion, and what evidence must accompany that declaration. The architecture proposed here makes that runtime contract explicit through five planes, packetized state, and verify-gated completion in this bounded case-study setting, together with fail-closed semantics, bounded recovery, supervised rollback, trace-first observability, and a learning loop that stays off the hot path.

Within those limits, the support is architectural and methodological rather than performance-oriented. A read-only verify gate plus packetized state make completion decisions inspectable and easier to withhold when evidence is weak. The released empirical slice is synthetic-heavy, concentrated, and small on production-classified events, so the results should be read as operational evidence about the control path rather than as broad performance estimates. For implementers, the implication is conservative: keep PGV advisory, keep verify read-only, and tie completion to packetized evidence rather than to free-form worker assertions.

Stronger empirical validation would require aligned task-level denominators, larger non-synthetic corpora, better blocked-reason instrumentation, true latency and token accounting, and controlled gated-versus-ungated ablations. The first practical step is a stable join between task-summary and event ledgers so that a completion-eligible task denominator can be defined and reproduced. A second practical step is sharper blocked-state instrumentation, including fields such as \path{blocked_reason_class}, \path{blocking_predicate}, \path{missing_packet_type}, \path{evidence_floor_failed}, \path{owner_gap}, and \path{stale_ground}. These additions would move the work from a bounded case-study report toward stronger empirical validation.

\section*{CRediT authorship contribution statement}

\textbf{Hai-Duong Nguyen:} Conceptualization, Methodology, Formal analysis, Writing (original draft), Writing (review and editing). \textbf{The-Xuan Tran:} Conceptualization, Writing (review and editing). Both authors reviewed and approved the final manuscript.

\section*{Declaration of competing interest}

The authors declare that they have no known competing financial interests or personal relationships that could have appeared to influence the work reported in this paper.

\section*{Funding}

The authors received no specific grant from any funding agency in the public, commercial, or not-for-profit sectors.

\section*{Data availability}

The raw runtime traces underlying this case study are not publicly released because they contain operator-specific metadata and internal project traces. Publicly released materials are limited to Supplementary Information~1 (ESM\_1.pdf), which contains a tabulated synthetic event sample, and a compact supplementary accounting and schema package containing an anonymized schema file, minimal verify-gate pseudocode, packet templates, denominator-provenance notes, subset task-accounting notes, concentration notes, overhead-proxy notes, event-taxonomy-parity notes, and aggregate verify and PGV breakdowns. A file-level manifest is provided in the supplementary accounting and schema package. Claims described as supported or checkable in this manuscript refer to those supplied supplementary materials; without them, the manuscript alone provides the narrative summary but not independent reproduction of the aggregate accounting. These materials let readers inspect the accounting logic and paper-facing schema semantics, but they do not reproduce the full internal event corpus, the restricted internal audits, or the internal orchestration environment. Additional anonymized aggregate summaries and representative anonymized traces may be shared with editors or reviewers under reasonable confidentiality constraints if journal process permits.

\section*{Code availability}

No public repository for the internal reference implementation accompanies this manuscript. The supplementary accounting and schema package includes schema definitions and documentation notes, but not the internal orchestration code, private prompts, or the private operating environment.

\section*{Ethics approval}

Not applicable.

\section*{Consent to participate}

Not applicable.

\section*{Consent for publication}

Not applicable.

\section*{Declaration of generative AI and AI-assisted technologies in the writing process}

During preparation of this work, the authors used generative AI tools for language-level review, structural feedback, and internal consistency checking. After using these tools, the authors reviewed, edited, and verified all content and take full responsibility for the final manuscript. The multi-agent system studied in this paper was not used to generate the reported data or empirical results.

\section*{Acknowledgements}

No external acknowledgements are declared.

\begingroup
\footnotesize
\RaggedRight%
\sloppy
\setlength{\emergencystretch}{3em}

\endgroup


\begin{thebibliography}{99}
\setlength{\itemsep}{0.25em}
\setlength{\parskip}{0pt}

\bibitem{react}
S.~Yao, J.~Zhao, D.~Yu, N.~Du, I.~Shafran, K.~R. Narasimhan, and Y.~Cao, ``ReAct:\@ Synergizing reasoning and acting in language models'', in \emph{International Conference on Learning Representations (ICLR)}, 2023. [Online]. Available: \url{https://openreview.net/forum?id=WE_vluYUL-X}.

\bibitem{camel}
G.~Li, A.~Hammoud, B.~Itani, D.~Khizbullin, and B.~Ghanem, ``{CAMEL}:\@ Communicative agents for mind exploration of large language model society'', \emph{arXiv preprint arXiv:2303.17760}, 2023. DOI:\@ \url{https://doi.org/10.48550/arXiv.2303.17760}.

\bibitem{autogen}
Q.~Wu, G.~Bansal, J.~Zhang, Y.~Wu, B.~Li, E.~Zhu, L.~Jiang, X.~Zhang, S.~Zhang, J.~Liu, A.~H. Awadallah, R.~W. White, D.~Burger, and C.~Wang, ``AutoGen:\@ Enabling next-gen LLM applications via multi-agent conversations'', in \emph{Conference on Language Modeling (COLM)}, 2024. [Online]. Available: \url{https://openreview.net/forum?id=BAakY1hNKS}.

\bibitem{metagpt}
S.~Hong, M.~Zhuge, J.~Chen, X.~Zheng, Y.~Cheng, J.~Wang, C.~Zhang, Z.~Wang, S.~K. S. Yau, Z.~Lin, L.~Zhou, C.~Ran, L.~Xiao, C.~Wu, and J.~Schmidhuber, ``MetaGPT:\@ Meta programming for a multi-agent collaborative framework'', in \emph{International Conference on Learning Representations (ICLR)}, 2024. [Online]. Available: \url{https://openreview.net/forum?id=VtmBAGCN7o}.

\bibitem{reflexion}
N.~Shinn, F.~Cassano, A.~Berman, and A.~Goyal, ``Reflexion:\@ Language agents with verbal reinforcement learning'', \emph{arXiv preprint arXiv:2303.11366}, 2023. DOI:\@ \url{https://doi.org/10.48550/arXiv.2303.11366}.

\bibitem{selfrefine}
A.~Madaan, N.~Tandon, P.~Clark, Y.~Yang, M.~Faruqui, P.~Parikh, Y.~Zhang,\\
V.~Nangia, S.~Fried, and I.~Celikyilmaz, ``Self-Refine:\@ Iterative refinement with self-feedback'', \emph{arXiv preprint arXiv:2303.17651}, 2023. DOI:\@ \url{https://doi.org/10.48550/arXiv.2303.17651}.

\bibitem{sweagent}
J.~Yang, C.~E. Jimenez, A.~Wettig, K.~Lieret, S.~Yao, K.~R. Narasimhan, and O.~Press, ``SWE-agent: Agent-computer interfaces enable automated software engineering'', in \emph{Advances in Neural Information Processing Systems (NeurIPS)}, 2024. [Online]. Available: \url{https://openreview.net/forum?id=mXpq6ut8J3}.

\bibitem{cove}
S.~Dhuliawala, M.~Karpinska, P.~R. Gribovskaya, V.~Stoyanov, and A.~Agha, ``Chain-of-Verification reduces hallucination in large language models'', \emph{arXiv preprint arXiv:2309.11495}, 2023. DOI:\@ \url{https://doi.org/10.48550/arXiv.2309.11495}.

\bibitem{constitutional}
Y.~Bai, A.~Jones, K.~Ndousse, A.~Askell, A.~Chen, N.~DasSarma, D.~Drain, S.~Ganguli, T.~Henighan, N.~Joseph, B.~Mann, A.~Olsson, C.~Olsson, B.~Pursell, J.~Skalse, E.~Perez, and J.~Kaplan, ``Constitutional AI:\@ Harmlessness from AI feedback'', \emph{arXiv preprint arXiv:2212.08073}, 2022. DOI:\@ \url{https://doi.org/10.48550/arXiv.2212.08073}.

\bibitem{memgpt}
C.~Packer, S.~Fang, K.~Patel, Y.~Lin, S.~Wooders, and V.~K. Kuleshov, ``MemGPT:\@ Towards LLMs as operating systems'', \emph{arXiv preprint arXiv:2310.08560}, 2023. DOI:\@ \url{https://doi.org/10.48550/arXiv.2310.08560}.

\bibitem{agentbench}
X.~Liu, H.~Yu, H.~Zhang, Y.~Xu, X.~Lei, H.~Lai, Y.~Gu, H.~Ding, K.~Men, K.~Yang, and others, ``AgentBench:\@ Evaluating LLMs as agents'', in \emph{International Conference on Learning Representations (ICLR)}, 2024. [Online]. Available: \url{https://openreview.net/forum?id=zAdUB0aCTQ}.

\bibitem{gaia}
G.~Mialon, C.~Fourrier, T.~Wolf, Y.~LeCun, and T.~Scialom, ``{GAIA}:\@ a benchmark for General AI Assistants'', in \emph{International Conference on Learning Representations (ICLR)}, 2024. [Online]. Available: \url{https://openreview.net/forum?id=fibxvahvs3}.

\bibitem{llmjudge}
L.~Zheng, W.-L. Chiang, Y.~Sheng, S.~Zhuang, Z.~Wu, Y.~Zhuang, Z.~Lin, Z.~Li, D.~Li, E.~P. Xing, H.~Zhang, J.~E. Gonzalez, and I.~Stoica, ``Judging LLM-as-a-judge with MT-Bench and Chatbot Arena'', \emph{Advances in Neural Information Processing Systems}, vol.~36, 2023, Datasets and Benchmarks Track. [Online]. Available: \url{https://proceedings.neurips.cc/paper_files/paper/2023/hash/91f18a1287b398d378ef22505bf41832-Abstract-Datasets_and_Benchmarks.html}.

\bibitem{llmevalbias}
R.~Stureborg, D.~Alikaniotis, and Y.~Suhara, ``Large language models are inconsistent and biased evaluators'', \emph{arXiv preprint arXiv:2405.01724}, 2024. DOI:\@ \url{https://doi.org/10.48550/arXiv.2405.01724}.

\bibitem{langgraph}
LangChain, ``LangGraph overview'', online documentation. Available: \url{https://docs.langchain.com/oss/python/langgraph/overview}.

\bibitem{crewai}
CrewAI, ``Introduction'', online documentation. Available: \url{https://docs.crewai.com/en/introduction}.

\bibitem{openaiguardrails}
OpenAI, ``Guardrails and human review'', online documentation. Available: \url{https://developers.openai.com/api/docs/guides/agents/guardrails-approvals}.

\bibitem{openaihitl}
OpenAI, ``Running agents'', online documentation. Available: \url{https://developers.openai.com/api/docs/guides/agents/running-agents}.

\bibitem{semantickernel}
Microsoft, ``Semantic Kernel agent framework'', online documentation. Available: \url{https://learn.microsoft.com/en-us/semantic-kernel/frameworks/agent/}.

\bibitem{dapper}
B.~H. Sigelman, L.~A. Barroso, M.~Burrows, P.~Stephenson, M.~Plakal, D.~Beaver, S.~Jaspan, and C.~Shanbhag, ``Dapper, a large-scale distributed systems tracing infrastructure'', Google Research, technical report, 2010.

\bibitem{humblecd}
J.~Humble and D.~Farley, \emph{Continuous Delivery: Reliable Software Releases through Build, Test, and Deployment Automation}. Boston, MA, USA:\@ Addison-Wesley, 2010.

\bibitem{accelerate}
N.~Forsgren, J.~Humble, and G.~Kim, \emph{Accelerate: The Science of Lean Software and DevOps}. Portland, OR, USA:\@ IT Revolution, 2018.

\bibitem{weillross}
P.~Weill and J.~W. Ross, \emph{IT Governance: How Top Performers Manage IT Decision Rights for Superior Results}. Boston, MA, USA:\@ Harvard Business School Press, 2004.

\end{thebibliography}
\end{document}